\documentstyle[12pt]{article}
\begin{document}


\title{Irreducibility of $n$-ary quantum information}
\author{Karl Svozil\\
 {\small Institut f\"ur Theoretische Physik, University of Technology Vienna }     \\
  {\small Wiedner Hauptstra\ss e 8-10/136,}
  {\small A-1040 Vienna, Austria   }            \\
  {\small e-mail: svozil@tuwien.ac.at}}
\date{ }
\maketitle

\begin{abstract}
Quantum information
is radically different from classical information
in that the quantum formalism (Hilbert space)
makes necessary the introduction of irreducible ``nits,''
$n$ being an arbitrary natural number (bigger than one);
not just bits.
\end{abstract}

As pointed out many times by Landauer and others
(e.g., \cite{landauer,feynman-computation})
the formal concept of {\em information} is tied
to physics,
at least as far as applicability is a concern.
Thus it should come as no surprise that quantum mechanics
requires fundamentally new concepts of information
as compared to the ones appropriate for classical physics.
And indeed, research into quantum information and computation theory
has exploded in the last decade, bringing about a wealth of new
ideas and formalisms.

There seems to be one issue,
which, despite notable exceptions (e.g., \cite[Footnote 6]{zeil-99}),
has not yet been acknowledged widely:
the principal irreducibility of $n$-ary quantum information
associated with the $n$-dimensionality of Hilbert space.
A physical configuration allowing for $n$ possible outcomes
has to be encoded quantum mechanically by an $n$-dimensional Hilbert space.
Any single one of the $n$ basis vector corresponds
to a onedimensional subspace spanned by that basis vector
which in turn corresponds to the following physical proposition:
{\em ``the physical system is in a pure state corresponding to the basis vector.''}
In more operational terms, a particle can be prepared
in a single one of $n$ possible states.
This particle then carries the information to
``be in a single one from $n$ different states.''
Subsequent measurements may confirm this statement.
Thus the most natural code basis for this configutaion is an $n$-ary code,
and not a binary one.

Classically, there is no preferred code basis whatsoever.
Because every classical state is postulated to be determined by a
point in phase space.
Formally, this amounts to an infinite amount information in whatever base,
since
with probability one, all points are random; i.e., algorithmically incompressible.
Operationally, only a finite amount of classical information is accessible.
Yet, in what base this finite amount of classical information is coded
is purely conventional and depends on the particular choice of the experimental setup.

Note that
by a well known theorem for unitary operators
\cite{murnaghan},
any quantum measurement of an $n$-ary system can be decomposed
into binary measurements.
Also, it is possible to group the $n$ possible outcomes into binary filters of
ever finer resultion; calling the successive outcomes of these filter process
the ``binary code.''
Yet, all these attempts result in codes with undesirable features.
Unitary decompositions in general yield non-comeasurable observables and thus to
non-operationalizability. Filters are inefficient, and so may be binary codes
\cite{Cal-Cam-96}.

So far, the main emphasis in the area of quantum computation
has been in the area of binary decision problems.
It is suggested that these investigations should be extended to
$n$-ary decision problems (e.g., \cite[pp. 332-340]{kleene-52}),
for which quantum information theory seems
to be extraordinarily well equipped.


\end{document}